\documentclass[%
 reprint,
nofootinbib,
 amsmath,amssymb,
 aps,
]{revtex4-1}

\usepackage[dvipdfmx]{graphicx}
\usepackage{dcolumn}
\usepackage{bm}
\usepackage{subfigure}
\usepackage{amsmath}

\usepackage{ulem,color}
\usepackage{hyperref}

\begin{document}

\title{Toward a simultaneous resolution of the $H_0$ and $S_8$ tensions: early dark energy and an interacting dark sector model}

\author{Mai Yashiki}
 \email{yashikimi@nbu.ac.jp}
\affiliation{Faculty of School of Engineering, Nippon Bunri University, Oita-shi, Oita 870-0397, Japan}

\date{\today}

\begin{abstract}
The tension between the Hubble constant ($H_0$) inferred from the cosmic microwave background (CMB) and that measured from late-time observations, such as the local distance ladder, is a major challenge in modern cosmology.
Early dark energy (EDE) has been proposed as a possible resolution to the $H_0$ tension, but it typically worsens the $S_8$ tension by enhancing the small-scale matter power spectrum due to an increased cold dark matter density.
To address this issue, we propose a model that combines EDE with an interacting dark energy-dark matter (iDEDM) scenario, and investigate whether this mixed model can simultaneously resolve both tensions.
We find that the DE-DM interaction suppress the growth of structure and reduce $S_8$, while EDE contributes to increase $H_0$, although less effectively than in the EDE-only case.
Our MCMC analysis using Planck 2018, DESI BAO, DES, Pantheon+, and SH0ES data shows that the mixed model provides modest improvements in both tensions, although it does not fully resolve either.
This limitation appears to stem from the fact that both EDE and iDEDM independently favor a higher present-day matter density, which reduces the angular diameter distance and limits the degree to which EDE can lower the sound horizon.
\end{abstract}

\pacs{Valid PACS appear here}

\maketitle

\section{\label{sec:intro}Introduction}

The $\Lambda$CDM model, which is the most standard cosmological model, is consistent with observations, including precise measurements of the cosmic microwave background (CMB).
However, the Hubble constant $H_0$, which is the current expansion rate of the universe, has shown a persistent discrepancy between estimates from the early and late universe.
This is called the "Hubble tension" and has become a serious problem (see~\cite{Knox2020} for a comprehensive review of the Hubble tension and possible theoretical solutions).
The latest observation of CMB from the Planck satellite, based on $\Lambda$CDM, yields $H_0 = 67.4\pm0.5$ km/s/Mpc~\cite{planck18}, while the local distance ladder measurement of $H_0$ via Cepheid-calibrated SNeIa from SH0ES yields $H_0 = 74.03 \pm 1.42$ km/s/Mpc~\cite{sh0es}.
This discrepancy is about $5\sigma$, indicating a statistically significant tension.
Moreover, the late-time observation using the gravitational lensing, whose measurement is independent of the distance ladder, gives almost the same value as SH0ES: $H_0 = 73.3^{+1.7}_{-1.8}$~\cite{h0licow}, and several other local observations are also consistent with the higher value of $H_0$~\cite{LIGOH0, TRGBH0,HSTH0,TFH0,TRGBH02,TDCOSMOH0,2308.01875,2311.13062,JWSTH0}.
As systematic errors in observations cannot completely explain this inconsistency, the Hubble tension may imply the new physics beyond $\Lambda$CDM model (of course, many studies have investigated whether systematic errors in observations can explain the $H_0$ tension; see e.g.~\cite{Freedman2021,Mortsell2022,Mortsell20222,Riess2023,Riess2024,Breuval2024}).
There are two main approaches to modifying the $\Lambda$CDM model in order to resolve the Hubble tension: one is to modify the late-time universe, and the other is to modify the early universe.
Late-time models have been found to be less successful in resolving the Hubble tension, mainly because increasing $H_0$ without modifying the sound horizon at baryon drag, $r_d$, leads to tensions with BAO and Hubble flow SNeIa measurements~\cite{Vagnozzi2020,Alesta2020,Banihashemi2021,Heisenberg2022,HEISENBERG2023,Gao2024,Yao2024}.

One solution to the Hubble tension by modifying the early universe\footnote{Several models have been proposed to resolve the Hubble tension by modifying the early universe, such as the EDE scenario; see e.g.~\cite{Sekiguchi-Takahashi, Niedermann2021, Karwal2022, Lin2019, Odintsov2023, Sakstein2020}. For a comprehensive review, see Ref.~\cite{Poulin2023-review}.} is early dark energy (EDE)~\cite{ede-karwal, ede-poulin, Braglia2020, Forconi2024}.
This scenario introduces a new scalar field component that behaves like a cosmological constant around the the matter-radiation equality, inducing a brief period of accelerated expansion, and subsequently decays faster than matter after the matter-radiation equality.
Since EDE modifies the expansion rate in the early universe, the comoving sound horizon at decoupling, $r_s$, becomes smaller, which keeps the angular size of the sound horizon, $\theta_s = r_s/D_A$, unchanged.
Here, $\theta_s$ is tightly constrained by CMB observations from Planck~\cite{planck18}, and the angular diameter distance $D_A$ scales approximately as $1/H_0$.
To resolve the Hubble tension, the energy fraction of EDE near recombination must be around 10\%.
Previous research~\cite{ede-smith} reported $H_0 = 72.19\pm 1.20$ km/s/Mpc using a combination of Planck, baryon acoustic oscillation (BAO), redshift-space distortion (RSD), and SH0ES data, indicating that EDE can successfully reconcile the tension.
In a recent study~\cite{2505.08051}, based on  Planck low-$\ell$, ACT DR6,  Planck lensing, Pantheon+, and DESI DR2 data, the authors reported $f_{\rm EDE} = 0.09 \pm 0.03$ and $H_0 = 71.0 \pm 1.1$ km/s/Mpc without adopting the SH0ES prior.
When including the SH0ES prior, their analysis finds that the EDE model favors a higher value of $H_0 \sim 73$ km/s/Mpc.

Unfortunately, the resolutions involving EDE become less viable when large-scale structure (LSS) data are taken into account.
As $H_0$ increases due to the introduction of the EDE phase, other cosmological parameters must shift to maintain consistency with the CMB temperature and the polarization power spectrum~\cite{ede-Hill, ede-ivanov}.
In particular, the cold dark matter density parameter, $\omega_{\rm cdm}$, tends to increase, which in turn raises the amplitude of matter fluctuations quantified by $S_8 \equiv \sigma_8 (0.3/\Omega_m)^{1/2}$, where $\sigma_8$ is the RMS mass fluctuation within spheres of radius 8 Mpc/$h$ at $z=0$.
This exacerbates the so-called $S_8$ tension, a 2--3$\sigma$ discrepancy between CMB and LSS measurements~\cite{S8tension1, S8tension2}.
The Planck result based on $\Lambda$CDM gives $S_8 = 0.830 \pm 0.013$~\cite{planck18}, while LSS observations yield lower values, such as $S_8 = 0.745\pm 0.039$ from KiDS~\cite{kids} and $S_8 = 0.759^{+0.024}_{-0.021}$ from DES-Y3~\cite{des-y3}.
In the EDE model, previous work~\cite{ede-smith} obtained $S_8 = 0.842 \pm 0.014$ (without using LSS data), which worsens the tension compared to $\Lambda$CDM.
When including the BOSS DR12 data~\cite{boss}, the inferred value of the Hubble constant drops to $H_0 = 68.73^{+0.42}_{-0.69}$ km/s/Mpc~\cite{ede-ivanov}, suggesting that EDE alone cannot fully resolve the Hubble tension.
Refs.~\cite{vagnozzi2023, Jedamzik2021} also argued that early-time new physics alone, such as EDE, is insufficient to fully resolve the $H_0$ tension (see the papers for details).

It has been pointed out that resolving the $H_0$ and $S_8$ tensions simultaneously requires modifications to both the early and late universe, see, e.g., Ref.~\cite{clark2021}.
Several studies have investigated whether combining early- and late-time modifications can simultaneously resolve both the $H_0$ and $S_8$ tensions~\cite{2407.01173toda,2302.07333,2503.21600}.
However, none of these models has succeeded in fully resolving both tensions\footnote{See also~\cite{2502.14608} for a different approach combining EDE with additional physics.}.

In this paper, we focus on the suppression of the matter power spectrum on small scales in order as means to alleviate the $S_8$ tension.
Interactions between dark energy (DE) and dark matter (DM), known as interacting DE-DM models (iDEDM)\footnote{For original works on interacting models, see~\cite{0901.1611}; for reviews, see~\cite{Wang2016,2402.00819}.}, have been investigated as a potential solution to the $S_8$ tension~\cite{Lucca2021,Sabogal2024,Shah2024,2502.03390,2503.23225}.
These phenomenological models modify the background and perturbation evolution by introducing energy-momentum exchange between DE and DM, without relying on an underlying Lagrangian structure\footnote{While this work focuses on phenomenological models, there also exist studies that derive DE-DM interactions from underlying particle physics frameworks; see, e.g.,~\cite{2406.19284} for a Lagrangian-based approach.}.
The sign of the interaction determines the direction of energy transfer: in the iDEDM scenario, energy flows from DE to DM.
This leads to a suppressed DM density at early times relative to $\Lambda$CDM, which suppresses the growth of structure and hence reduces the amplitude of the matter power spectrum on small scales.
Ref.~\cite{Lucca2021} showed that iDEDM can lower the predicted $S_8$ to $0.802^{+0.015}_{-0.0112}$, while yielding $H_0 = 68.29^{+0.52}_{-0.46}$~km/s/Mpc.
This suggests that iDEDM can help alleviate the $S_8$ tension without significantly exacerbating the $H_0$ tension.

Since EDE and iDEDM affect the matter power spectrum in opposite ways---with EDE typically enhancing the amplitude and iDEDM suppressing it---it is natural to explore whether their combination can lead to a more successful resolution of both $H_0$ and $S_8$ tensions.
While EDE alone tends to worsen the $S_8$ tension despite increasing $H_0$, and iDEDM alone lowers $S_8$ but does not significantly increase $H_0$, their opposite effects may complement each other.
In this paper, we propose and investigate a unified model that incorporates both EDE and iDEDM, and examine whether this mixed model can simultaneously alleviate the $H_0$ and $S_8$ tensions.

The outline of this paper is as follows: in Sec.~\ref{sec:EDE}, we review the physics of the EDE model and the impact on the large-scale structure in detail.
In Sec.~\ref{sec:idedm}, we describe the mathematical setup of the iDEDM model considered in this work.
In Sec.~\ref{sec:mix}, we introduce our unified model that combines EDE and iDEDM.
In Sec.~\ref{sec:data}, we describe the datasets and analysis methods used in our parameter inference.
We present our main results in Sec.~\ref{sec:result}, and we conclude in Sec.~\ref{sec:conclusion}.

In the following, we adopt natural units where $8\pi G \equiv M_{\rm pl}^{-2} =1$, with $G$ being the gravitational constant and $M_{\rm pl}$ is the reduced Planck mass.

\section{\label{sec:EDE} Early Dark Energy}

We briefly review the EDE model and its impact on large-scale structure (LSS).
The background dynamics of EDE is governed by a canonical scalar field~\cite{ede-poulin,ede-smith,1409.0549}:
\begin{equation}
    \ddot{\phi} + 3H\dot{\phi} + \frac{dV(\phi)}{d\phi} = 0,
    \label{eq:EDE_bg}
\end{equation}
where a dot denotes a derivative with respect to cosmic time, and $H$ is the Hubble parameter.
The scalar field potential is given by
\begin{equation}
    V(\phi)=m^2f^2(1-\cos \phi/f)^n,
    \label{eq:EDE_potential}
\end{equation}
inspired by the ultra-light axion (ULA) scenario from string theory, where $m$ and $f$ are the axion mass and decay constant, respectively, and $n$ is a constant.
If the potential is sufficiently flat, the scalar field is initially frozen due to the Hubble friction term, $3H\dot{\phi}$, and behaves like a dark energy component.

To resolve the Hubble tension, the energy density fraction of EDE
\begin{equation}
    f_{\rm EDE} \equiv \frac{\rho_{\phi}}{\rho_{\rm tot}},
    \label{eq:f_ede}
\end{equation}
needs to reach approximately 10\% at the critical redshift $z_c$, defined by the condition $H(z_c) \simeq m$.
Here, $\rho_{\phi}$ and $\rho_{\rm tot}$ are the energy density of the EDE field and the total energy density of the universe, respectively.
As the universe expands and $H$ decreases, the scalar field is free from the Hubble friction and begins to roll down the minimum of the potential.
Around the potential minimum, Eq.~(\ref{eq:EDE_potential}) can be approximated as $V(\phi) \propto \phi^{2n}$, leading the field to oscillate.
The decay rate of the EDE field is then controlled by its effective equation of state as $\rho_{\phi}(t) \propto a(t)^{-3(1+w_{\phi})}$, which depends on the shape of the potential near the minimum.
The equation of state of EDE is given by
\begin{equation}
w_{\phi} = \left\{
\begin{array}{ll}
-1 & (z > z_c)\\
\frac{n-1}{n+1} & (z \leq z_c )
\end{array}
\right.,
\label{eq:w_ede}
\end{equation}
which shows that for $n=3$, the EDE field dilutes faster than radiation.

For the case of $n=3$, a recent analysis incorporating CMB, CMB lensing, BAO, RSD, SNIa, and SH0ES datasets found best-fit values of $H_0 = 72.19 \pm 1.20$ km/s/Mpc and $f_{\rm EDE} = 0.122 ^{+0.035}_{-0.030}$~\cite{ede-smith}.

\subsection{\label{sec:EDE_LSS} EDE meets LSS}

\begin{figure}
    \centering
    \includegraphics[width=8cm]{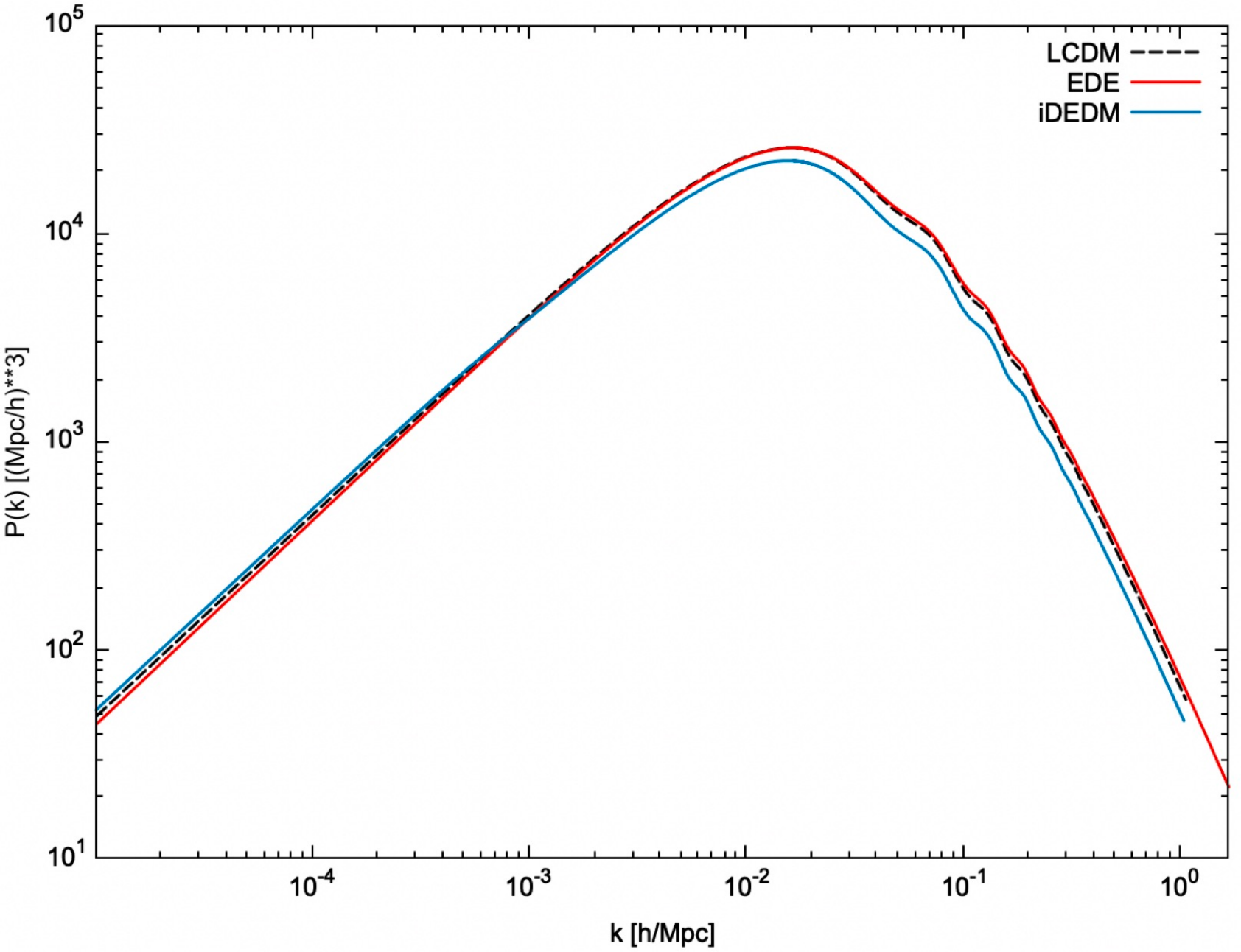}
    \caption{Non-linear matter power spectrum $P(k)$ at $z = 0$ for $\Lambda$CDM (dashed black line), EDE (solid red line) and iDEDM (solid blue line) models. We use the best-fit values from Ref.~\cite{ede-smith} for EDE and Ref.~\cite{Lucca2021} for iDEDM.}
    \label{fig:mPS_separate}
\end{figure}

As mentioned above, while the EDE model can achieve a high value of $H_0$, it no longer provides a good fit to the CMB power spectra unless other cosmological parameters are adjusted.
In particular, the cold dark matter density parameter, $\omega_{\rm cdm} = \Omega_{\rm cdm}h^2$, where $h$ is the reduced Hubble parameter, must increase in order to maintain the fit to the CMB temperature power spectrum.
An increased $\omega_{\rm cdm}$ enhances the matter density at early times, which deepens gravitational potentials and accelerates the growth of structure.
This leads to an enhancement of the small-scale matter power spectrum and, consequently, to an increase in $S_8$, which characterize the amplitude of matter fluctuations.
Indeed, Ref.~\cite{ede-smith} reported $\omega_{\rm cdm} = 0.1306$, about 10\% higher than the $\Lambda$CDM value, and $S_8 = 0.842 \pm 0.014$, which worsens the $S_8$ tension.
Fig.~\ref{fig:mPS_separate} shows the non-linear matter power spectrum at $z=0$ for $\Lambda$CDM, EDE, and iDEDM models, computed using the modified \texttt{CLASS} code for EDE~\cite{ede-Hill}, with the best-fit parameters from Ref.~\cite{ede-smith}.
As seen in the figure, the EDE model yields a higher amplitude of $P(k)$ at small scales compared to $\Lambda$CDM\footnote{Ref.~\cite{ede-Hill} attributed this increase primarily to shifts in the standard cosmological parameters, rather than the direct effect of the EDE field itself.}.
Since LSS data tightly constrain both the shape and amplitude of the matter power spectrum, $f_{\rm EDE}$ becomes tightly limited when LSS data are included.
In Ref.~\cite{ede-Hill}, analysis using DES-Y1, KiDS, and HSC datasets gave $H_0 = 68.75 \pm 0.50$ km/s/Mpc and $f_{\rm EDE} < 0.053$, without the SH0ES prior.
These findings suggests that the EDE model alone cannot resolve the $H_0$ tension while remaining consistent with LSS data.

To address this issue, Ref.~\cite{2302.07333} explored whether modifying the late-time expansion history could suppress the $S_8$ enhancement induced by EDE.
They introduced a redshift-binned dynamical dark energy model that allows $\Omega_m$ to decrease at late times.
Although their model achieved $H_0 = 70.6$, $f_{\rm EDE} = 0.064$, and $S_8 = 0.819$ even with LSS data, the improvement was insufficient to resolve both tensions simultaneously.

We are therefore motivated to explore an alternative approach: suppressing the small-scale matter power spectrum directly, in order to lower $S_8$ without compromising the EDE-induced increase in $H_0$.
Since $S_8$ depends on both $\Omega_m$ and $\sigma_8$, and the latter is primarily determined by the amplitude of the matter power spectrum in the range $0.1 h/{\rm Mpc} \lesssim k \lesssim 1 h/{\rm Mpc}$, controlling $P(k)$ at small scales becomes crucial.
As $\omega_{\rm cdm}$ increases (decreases), the small-scale amplitude of $P(k)$ also increases (decreases), directly impacting $\sigma_8$.
In this work, we focus on an interacting dark energy-dark matter (iDEDM) scenario as a mechanism to suppress small-scale power and reduce $S_8$, while retaining the early-time dynamics of EDE.

\section{\label{sec:idedm} Interacting dark sector model}

We briefly review the interacting dark sector model and its mathematical setup.
In this paper, we focus on the phenomenological interaction described at the level of fluid dynamics (see~\cite{0901.1611} for more details).

At the background level, dark energy (DE) and dark matter (DM) are not conserved independently but are coupled via an energy transfer function $Q$,
\begin{align}
\dot{\rho}_{\rm c} + 3 H \rho_{\rm c} &= Q, \label{eq:cdm_energy_conservation} \\
\dot{\rho}_{\rm de} + 3 H \rho_{\rm de}(1+w_{\rm de}) &= -Q. \label{eq:de_energy_conservation}
\end{align}
where $\rho_{\rm de}$ denotes the energy density of DE, and $w_{\rm de}$ is the equation of state of DE.
Here, $\rho_{\rm c}$ refers to the energy density of the cold dark matter, including the interacting component in the iDEDM model.
All other components, such as baryons and radiation, are assumed to be independently converved as usual, i.e., $\dot{\rho}_{\rm b} + 3H \rho_{\rm b} = 0$ and $\dot{\rho}_{\rm r} + 4H \rho_{\rm r} = 0$, where $\rho_{\rm b}$ and $\rho_{\rm r}$ denote the energy density of baryons and radiation.

We adopt a commonly used form of energy transfer, proportional to the DE energy density\footnote{Alternative forms of $Q$ have also been proposed; see, e.g.,~\cite{2209.14816,2408.12403,2501.10323,2407.14934,2308.05807,2501.07361}.}:
\begin{equation}
    Q = \xi H \rho_{\rm de}.
    \label{eq:Q_1}
\end{equation}
Here, $\xi$ is the dimensionless coupling parameter that controls the interaction strength and direction.
For $\xi > 0$, energy flows from DE to DM, while $\xi < 0$ corresponds to the opposite energy flow.
In this paper, we refer to the $\xi>0$ case as the interacting DE-DM (iDEDM) model, and adopt this terminology throughout.
The iDEDM model has been proposed as a resolution to the $S_8$ tension, while the opposite scenario is often considered in the context of the $H_0$ tension~\cite{Gao2021,Nunes2022,Zhai2023,2404.15232,Forconi2024}.

In this interaction, the effective equations of state are given by~\cite{0901.1611}
\begin{align}
w^{\rm eff}_{\rm de} &= w_{\rm de} + \frac{\xi}{3}, \label{eq:w_de-const} \\
w^{\rm eff}_{\rm c} &= -\frac{\xi}{3}. \label{eq:w_dm}
\end{align}
Here, $w^{\rm eff}_{\rm de}$ and $w^{\rm eff}_{\rm c}$ represent the effective equations of state for dark energy and cold dark matter, respectively, reflecting how the interaction modifies their background evolution.
In this formulation, particular care must be taken in the choice of $w_{\rm de}$ to avoid gravitational and early-time instabilities~\cite{0901.1611}.
To ensure numerical stability while maintaining consistency with $\Lambda$CDM in the $\xi = 0$ limit, we follow the common approach of slightly shifting $w_{\rm de}$ to $-1.001$.

Fig.~\ref{fig:mPS_separate} shows the matter power spectrum for the iDEDM model, calculated using a modified version of the public \texttt{CLASS} code~\cite{LuccaclassiDMDE}\footnote{https://github.com/luccamatteo/class\_iDMDE}.
In the iDEDM case ($\xi > 0$), the DM density at early times is lower than in $\Lambda$CDM, which delays the matter-radiation equality.
This results in a leftward shift of the peak of the matter power spectrum.
The reduced early-time DM density also affects the decay rate of the gravitational potential, suppressing the growth of structure and lowering the amplitude of the power spectrum at small scales.
Additionally, since the DE-matter equality occurs later in iDEDM due to reduced early DM density, the duration of the matter perturbation growth is extended, leading to an enhancement of the matter power spectrum at large scales.
Thanks to these effects, this model suppresses the growth of matter fluctuations on small scales.
Ref.~\cite{Lucca2021} reported that, using Planck2018, BAO, Pantheon~\cite{pantheon}, KV450~\cite{KV450} and DES~\cite{des-y1,des-y13by2} data, the best-fit constraints are $\xi < 0.12$, $H_0 = 68.29^{+0.52}_{-0.46}$ km/s/Mpc and $S_8 = 0.802^{+0.015}_{-0.0112}$.
As a result, the iDEDM model can alleviate the $S_8$ tension without significantly worsening the $H_0$ tension.

\section{The Mixed model}
\label{sec:mix}
For all the reasons discussed above, we consider a mixed model that combines EDE and iDEDM, and investigate whether it can simultaneously resolve both the $H_0$ and $S_8$ tensions.
In this model, the scalar field $\phi$ represents the EDE field, while both DE and DM are treated as perfect fluids.
We do not consider any direct interaction between the EDE scalar field and DM.

\subsection{Background dynamics}
At the homogeneous and isotropic level, i.e., for the case of a Friedmann-Lema\^{i}tre-Roberston-Walker metric, the expansion rate of the universe can be written as
\begin{alignat}{2}
H^2       & = &\ & \frac{1}{3} \left( \rho_{\rm c} + \rho_{\rm de} + \rho_{\rm b} + \rho_{\rm r} + \rho_{\phi} \right), \\
\dot{H}   & = &\ & -\frac{1}{2} \Big( \rho_{\rm b} + \rho_{\rm c}(1+w^{\rm eff}_{\rm c}) + \frac{4}{3}\rho_{\rm r} \notag \\
          &   &\ & \phantom{-\frac{1}{2} \Big( \rho_{\rm b}} + \rho_{\rm de}(1+w^{\rm eff}_{\rm de}) + \rho_{\phi}(1+w_{\phi}) \Big),
\end{alignat}
where $w_\phi$, $w^{\rm eff}_{\rm de}$ and $w^{\rm eff}_{\rm c}$ are defined in Eqs.~(\ref{eq:w_ede}), (\ref{eq:w_de-const}) and (\ref{eq:w_dm}), respectively.
From Eqs.~(\ref{eq:cdm_energy_conservation}), (\ref{eq:de_energy_conservation}) and (\ref{eq:Q_1}), the energy densities of DE and DM are expressed as
\begin{align}
\rho_{\rm de} &= \rho_{\rm de,0}\, a^{-3(1+w^{\rm eff}_{\rm de})}, \\
\rho_{\rm c} &= \rho_{\rm c,0}\, a^{-3} + \frac{\xi}{3} \rho_{\rm de,0}\, a^{-3} \left( 1 - a^{-3w^{\rm eff}_{\rm de}} \right),
\end{align}
where $\rho_{\rm de,0}$ and $\rho_{\rm c,0}$ denote the present energy densities of DE and DM, respectively.
The energy density and pressure of the EDE filed are given by
\begin{align}
\rho_{\phi} &= \frac{1}{2} \dot{\phi}^2 + V(\phi), \\
p_{\phi}    &= \frac{1}{2} \dot{\phi}^2 - V(\phi).
\end{align}
For convenience, the equation of motion for the EDE field, Eq.~(\ref{eq:EDE_bg}), can be rewritten using a redefined field variable, $\Theta \equiv \phi/f$, as
\begin{equation}
    \ddot{\Theta}+3H\dot{\Theta} + \frac{1}{f^2}V_{,\phi} = 0,
    \label{eq:EDE_bg_re}
\end{equation}
where $V_{, \phi} = dV/d\phi$.

\subsection{Perturbed dynamics}

The perturbation equations for the EDE scalar field are given by~\cite{ede-smith}
\begin{align}
\delta_\phi' &= -(1+w_\phi)\left( \theta_\phi + \frac{1}{2}h' \right) -6\mathcal{H}\delta_\phi \notag \\
&\phantom{=} -9(1-c^2_\phi)(1+w_\phi)\mathcal{H}^2\frac{\theta_\phi}{k^2}, \\
\theta_\phi' &= 2\mathcal{H}\theta_\phi + \frac{\delta_\phi}{1+w_\phi},
\label{eq:EDE_perturb}
\end{align}
where $\delta_\phi$ and $\theta_\phi$ are the density contrast and velocity divergence of the EDE scalar field, respectively, and the prime denotes a derivative with respect to conformal time, and $\mathcal{H} = aH$ is the conformal Hubble parameter.
Here, $h$ is the synchronous gauge metric perturbation and $k$ is the comoving wave number.
The EDE field adiabatic sound speed is given by
\begin{equation}
    c^2_\phi \equiv \frac{\dot{p_\phi}}{\dot{\rho_\phi}} = 1 + \frac{2}{3}a^2\frac{V_{,\phi}}{\mathcal{H}^2\phi'}.
\end{equation}

The coupling between the two dark sectors will affect the evolution of the DE and DM density perturbations.
The perturbation equations for DE and DM in the synchronous gauge are given as~\cite{0901.1611, Lucca2021, 2404.06396}
\begin{align}
\delta'_{\rm de} &= - (1+w_{\rm de})\left(\theta_{\rm de} + \frac{h'}{2}\right) - \frac{h'}{6} \xi \notag \\
&\phantom{=} - 3 \mathcal{H} (1-w_{\rm de}) \Bigg[ \delta_{\rm de} + \frac{\mathcal{H}\theta_{\rm de}}{k^2} \left\{3(1+w_{\rm de})+\xi\right\} \Bigg], \\
\theta'_{\rm de} &= 2 \mathcal{H} \theta_{\rm de} \left[1 + \frac{\xi}{1+w_{\rm de}} \left(1 - \frac{\theta_{\rm c}}{2\theta_{\rm de}} \right) \right] + \frac{k^2}{1+w_{\rm de}} \delta_{\rm de}, \\
\delta'_{\rm c} &= \theta_{\rm c} -\frac{h'}{2} \left( 1- \frac{\xi}{3} \frac{\rho_{\rm de}}{\rho_{\rm c}} \right) + \xi \mathcal{H} \frac{\rho_{\rm de}}{\rho _{\rm c}} (\delta_{\rm de}-\delta_{\rm c}), \\
\theta'_{\rm c} &= - \mathcal{H} \theta_{\rm c}. 
\label{eq:iDEDM_perturb}
\end{align}
where $\delta_x$ and $\theta_x$ denote the density contrast and velocity divergence of DE ($x = \text{de}$) and DM ($x = \text{c}$), respectively.
As is commonly done in the literature, e.g.,~\cite{0901.1611}, we fix the DE rest-frame sound speed to unity, i.e., $c^2_{s, {\rm de}} = 1$, while the adiabatic sound speed of DE is $c^2_{a, {\rm de}}=w_{\rm de}$.

\section{Data}
\label{sec:data}
To evaluate the mixed model, we perform a Markov Chain Monte Carlo (MCMC) analysis using the public code \texttt{Cobaya}~\cite{cobaya,cobaya-desilike}, 
interfaced with our modified version of \texttt{CLASS}~\cite{CLASS}.
We consider a $6 + 4$ extension of the standard $\Lambda$CDM model with the following free parameters:
\[
\begin{array}{c}
    \{ H_0, \omega_{\rm b}, \omega_{\rm c}, n_s, \ln{(10^{10} A_s)}, \tau_{\rm reio} \} \\
   + \\
   \{\log_{10} (z_c), f_{\rm EDE}(z_c), \Theta_i, \xi\}.
\end{array}
\]
The first six parameters correspond to the standard $\Lambda$CDM cosmology.
The next three parameters describe the EDE sector---$\log_{10}(z_c)$, $f_{\rm EDE}(z_c)$, and $\Theta_i$ (the initial value of the rescaled EDE field $\Theta \equiv \phi/f$)---while the last one ($\xi$) is associated with the iDEDM model.
In this work, we fix $n = 3$ for the exponent of the scalar field potential defined in Eq.~(\ref{eq:EDE_potential}).
The prior distributions assumed in the MCMC analysis are summarized in Table~\ref{tab:prior}.
For most parameters, we adopt uniform (flat) priors.
For $\omega_{\rm b}$, $\tau_{\text{reio}}$, $\ln(10^{10} A_s)$, and $H_0$, we apply Gaussian priors informed by external measurements:
\begin{itemize}
  \item $\omega_{\rm b} = 0.02242 \pm 0.00049$ from BBN~\cite{BBNref}
  \item $\tau_{\text{reio}} = 0.054 \pm 0.007$ from Planck~\cite{planck18}
  \item $\ln(10^{10} A_s) = 3.044 \pm 0.014$ also from Planck~\cite{planck18}
  \item $H_0 = 73.2 \pm 1.3$ km/s/Mpc from SH0ES~\cite{sh0es}
\end{itemize}
We use the Gelman-Rubin convergence criterion~\cite{G-R}, stopping the chains when $|R - 1| < 0.01$ for the $\Lambda$CDM, EDE-only, and iDEDM-only models, while the chain for the mixed model was stopped at $|R - 1| = 0.02$\footnote{
The chain for the mixed model plateaued at $|R-1| \simeq 0.02$, while the posterior distributions and best-fit parameters remained stable. We have confirmed that this level of convergence does not significantly affect our main results.
}
We employ \texttt{GetDist}~\cite{GetDist} to analyze chains, obtain marginalized statistics, and plot confidence contours.
\begin{table}
    \centering
    \caption{Flat priors on the cosmological parameters used in the MCMC analysis.}
    \label{tab:prior}
    \begin{tabular}{lcc}
        \hline
        Parameter & Range \\
        \hline
        $\omega_{\rm c}$ &  $[0.001, 0.99]$ \\
        $n_s$ & $[0.8, 1.2]$ \\
        \hline
        $f_{\rm EDE}$ & $[0, 0.5]$ \\
        $\Theta_i$ & $[0.1, 3]$ \\
        $\log_{10}z_c$ &  $[2.8, 4]$ \\
        $\xi$ & $[0, 2]$ \\
        \hline
    \end{tabular}
\end{table}

In our MCMC analysis, we consider the following data sets:
\begin{itemize}
    \item {{\bf Planck CMB}}: We use the Planck 2018 baseline likelihood, which includes the low-$\ell$ temperature and polarization (low-$\ell$ TT and EE), the high-$\ell$ TT, TE, and EE spectra, and the CMB lensing reconstruction likelihood~\cite{planck18}. We refer to this data set as {\bf P18}.
    \item {{\bf DESI BAO}}: We use the DESI Year-1 (Y1) release BAO measurements~\cite{DESI-dr1}, based on galaxy clustering over $0.1 \leq z \leq 4.16$ in seven redshift bins including bright galaxy samples (BGS), luminous red galaxies (LRG), emission line galaxies (ELG), quasars (QSO), and the Lyman-$\alpha$ forest. We refer to this data set as {\bf DESI}.
    \item {{\bf DES}}: We adopt the joint 3$\times$2pt likelihood~\cite{des-y13by2} from the Dark Energy Survey Year 1 (DES Y1) results~\cite{des-y1}, combining cosmic shear, galaxy-galaxy lensing, and galaxy clustering measurements. 
    \item {{\bf Pantheon+}}: We use the Pantheon+ dataset~\cite{pantheonplus}, which includes 1701 Type Ia supernovae (SNIa), excluding the SH0ES calibration data. We refer to it as \textbf{PP}.
    \item \textbf{SH0ES}: We use the SH0ES measurement of the Hubble constant, $H_0 = 73.2 \pm 1.3$ km/s/Mpc~\cite{sh0es}, as a Gaussian prior. We refer to it as \textbf{H0}.
\end{itemize}

We first evaluate the quality of the fit using the minimum chi-squared value, $\chi^2_{\rm min}$, obtained for each model.
We also compute the difference in the minimum chi-squared values,
\begin{equation}
\Delta \chi^2 = {\chi^2_{\text{min}}}^{\Lambda\text{CDM}} - {\chi^2_{\text{min}}}^{\text{mixed}},
\end{equation}
to quantify the relative fit quality of the models without considering their complexity.
This allows us to assess the improvements in fit independently of the model's complexity.
A positive value of $\Delta \chi^2$ indicates that the mixed model provides a better fit to the data than the $\Lambda$CDM model.

In addition, since our model includes a large number of additional parameters, a lower value of $\chi^2$ alone does not necessarily imply a better fit to the observational data, as the improvement may simply reflect the increased flexibility of the model. 
To assess the overall performance of different models, we also compute the Akaike Information Criterion (AIC)~\cite{AIC1},
\begin{equation}
    {\rm AIC} = \chi^2_{\rm min} + 2k,
    \label{eq:AIC}
\end{equation}
where $\chi^2_{\rm min}$ is the minimum chi-squared value at the best-fit parameters, and $k$ is the number of free parameters in the model ($k=6$ for $\Lambda$CDM).
A lower AIC value indicates a better trade-off between goodness of fit and model complexity.
To compare the performance of our model with that of the $\Lambda$CDM model, we compute the difference in AIC values,
\begin{equation}
    \Delta {\rm AIC} = {\rm AIC}_{\Lambda {\rm CDM}} - {\rm AIC}_{\rm mixed},
    \label{eq:delta_AIC}
\end{equation}
following the standard interpretation~\cite{AIC2}, where $\Delta {\rm AIC} < 2$ indicates that both models are equally supported by the data, $2 \leq \Delta {\rm AIC} < 6$ suggests weak support for the mixed model, $6 \leq \Delta {\rm AIC} < 10$ suggests that the mixed model is disfavored, and $\Delta {\rm AIC} > 10$ indicates strong disfavor relative to $\Lambda$CDM.

\section{Results}
\label{sec:result}
\begin{figure*}
    \centering
    \includegraphics[width=18cm]{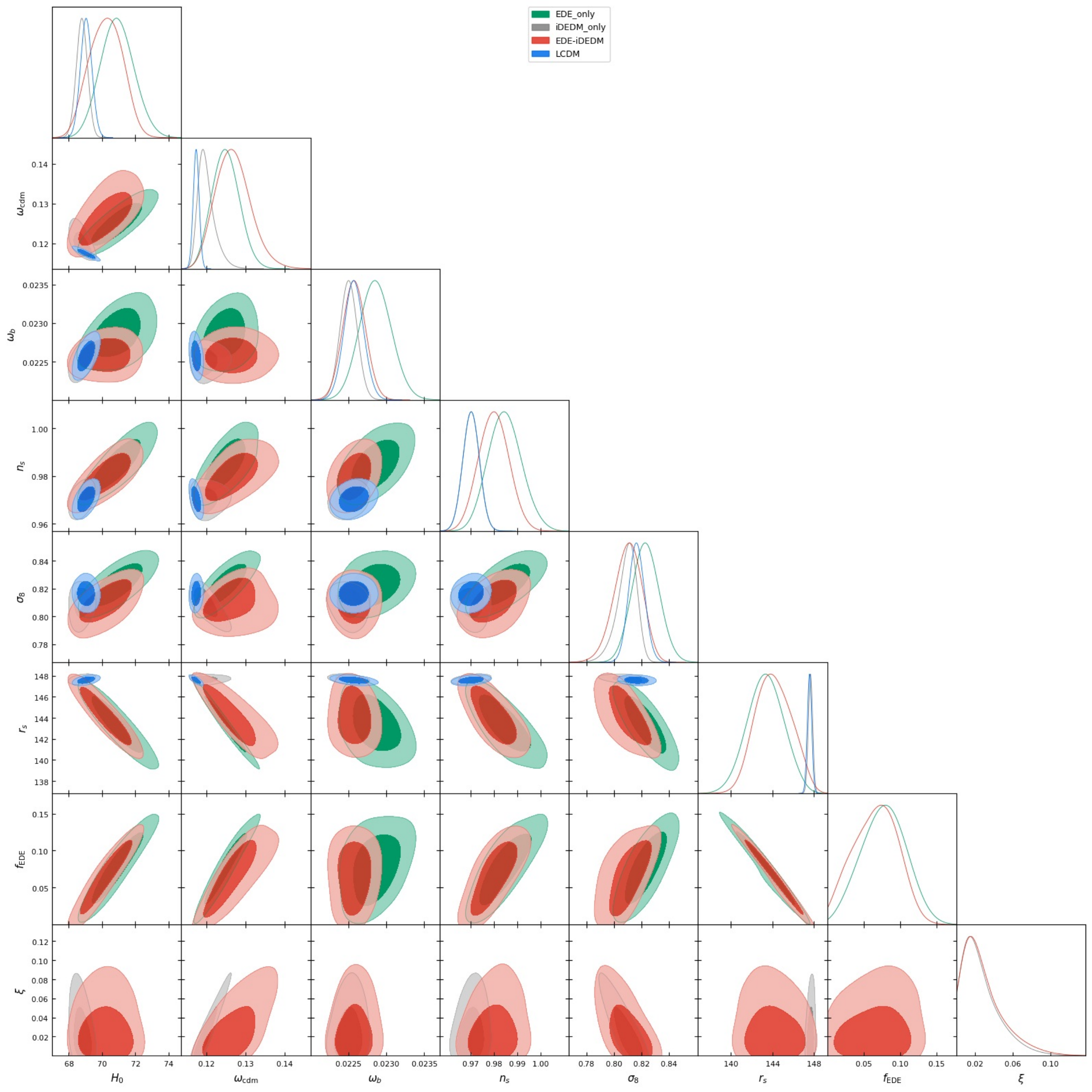}
    \caption{
    Cosmological parameter constraints from combined dataset of P18, DESI, DES, PP, and H0.
    The green, gray, red, and blue contours show $68\%$ and $95\%$ C.L. posteriors in the EDE-only, iDEDM-only, EDE-iDEDM (mixed), and $\Lambda$CDM models, respectively.
}
    \label{fig:triangle}
\end{figure*}

We now present the results of the MCMC analyses performed for our mixed model.
Fig.~\ref{fig:triangle} shows the triangle plot of selected cosmological parameters for our mixed model, $\Lambda$CDM, EDE-only, and iDEDM-only models, using all datasets described in the previous section.
The 68\% confidence level (C.L.) intervals for selected parameters are listed in Table~\ref{tab:result}.
In the EDE model, three parameters are sampled, but only the maximum fractional contribution, $f_{\rm EDE}$, is shown and reported, as it is the primary observable quantity; the other parameters primarily serve to set the initial conditions and are not directly constrained by the data.

\begin{table*}
    \centering
    \caption{Mean values and 68\% C.L. intervals of the cosmological parameters using the combined dataset of P18, DESI, DES, PP, and H0.}
    \label{tab:result}
    \begin{tabular}{|l|l|l|l|l|}
\hline
Parameter & $\Lambda$CDM & EDE & iDEDM & EDE-iDEDM\\
\hline

$H_0$ & $69.106\pm 0.352$ & $70.981\pm 0.892$ & $68.776\pm 0.329$ &  $70.000\pm 0.888$\\

$\omega_{\rm cdm}$ & $0.117\pm 0.001$ & $0.124\pm 0.003$ & $0.120\pm 0.001$ & $0.126\pm 0.003$\\

$n_s$ & $0.970\pm 0.004$ & $0.983\pm 0.006$ & $ 0.970\pm 0.003$ & $0.979\pm 0.006$\\

$\tau_{\rm reio}$ & $0.057\pm 0.007$ & $0.059\pm 0.007$ & $ 0.055\pm 0.004$ & $0.055\pm 0.004$\\

$\log A$ & $3.045\pm 0.014$ & $3.060\pm 0.014$ & $ 3.041\pm 0.008$ & $3.048 \pm 0.009$\\

$\Omega_m$ & $0.294\pm 0.004$ & $0.295\pm 0.004$ & $ 0.301\pm 0.005$ & $0.305\pm 0.005$ \\

$\sigma_8$ & $0.817\pm 0.006$ & $0.822\pm 0.010$ & $ 0.809\pm 0.008$ & $0.808\pm 0.010$\\

$r_s$ & $147.569 \pm 0.213$ & $143.673 \pm 1.709$ & $147.663 \pm 0.204$ & $144.523 \pm 1.950$\\
\hline

$f_{\rm EDE}$ & - &  $<0.128$ (95\% C.L.) &  - & $<0.113$ (95\% C.L.) \\

$\xi$ & - & - & $< 0.065$ (95\% C.L.) & $< 0.071$ (95\% C.L.) \\
\hline

$\Delta \chi ^2$ & - & $10.3097$ & $1.7781$ & $8.0840$ \\
$\Delta$ AIC & - & $4.3097$ & $-0.2219$ & $0.0840$ \\
\hline
    \end{tabular}
\end{table*}

As shown in Table~\ref{tab:result}, the mixed model yields a higher value of $H_0 = 70.000 \pm 0.888$ km/s/Mpc and a lower value of $\sigma_8 = 0.808 \pm 0.010$ compared to the $\Lambda$CDM model.
Using the definition of $S_8 = \sigma_8 \left(\Omega_m/0.3 \right)^{0.5}$, we obtain $S_8 = 0.815$ for the mixed model.
These results imply that, while the tensions are not fully resolved, the mixed model alleviates both the $H_0$ and $S_8$ tensions relative to $\Lambda$CDM.
The effect of iDEDM in the mixed model remains similar to that in the iDEDM-only case.
In contrast, the influence of EDE is noticeably diminished compared to its standalone case, suggesting that the presence of dark sector interactions limits the role EDE can play.
Altogether, the mixed model provides only a modest relaxation of the tensions, without fully resolving either.

Building on this, we examine the minimum $\chi^2$ and apply the AIC to evaluate model performance while accounting for the number of parameters.
As shown in Table~\ref{tab:result}, while the EDE-only model yields the most significant improvement in $\chi^2$, the mixed model also achieves a lower $\chi^2$ than $\Lambda$CDM.
However, after penalizing for the number of additional parameters via the AIC, the improvement becomes marginal: the mixed model shows only a slight preference over $\Lambda$CDM ($\Delta$AIC = 0.08), whereas the iDEDM-only model is mildly disfavored.
These results indicate that, despite a better fit in terms of $\chi^2$, the mixed model is not significantly favored over $\Lambda$CDM when accounting for model complexity.

\begin{figure}[h]
    \centering
    \includegraphics[width=8cm]{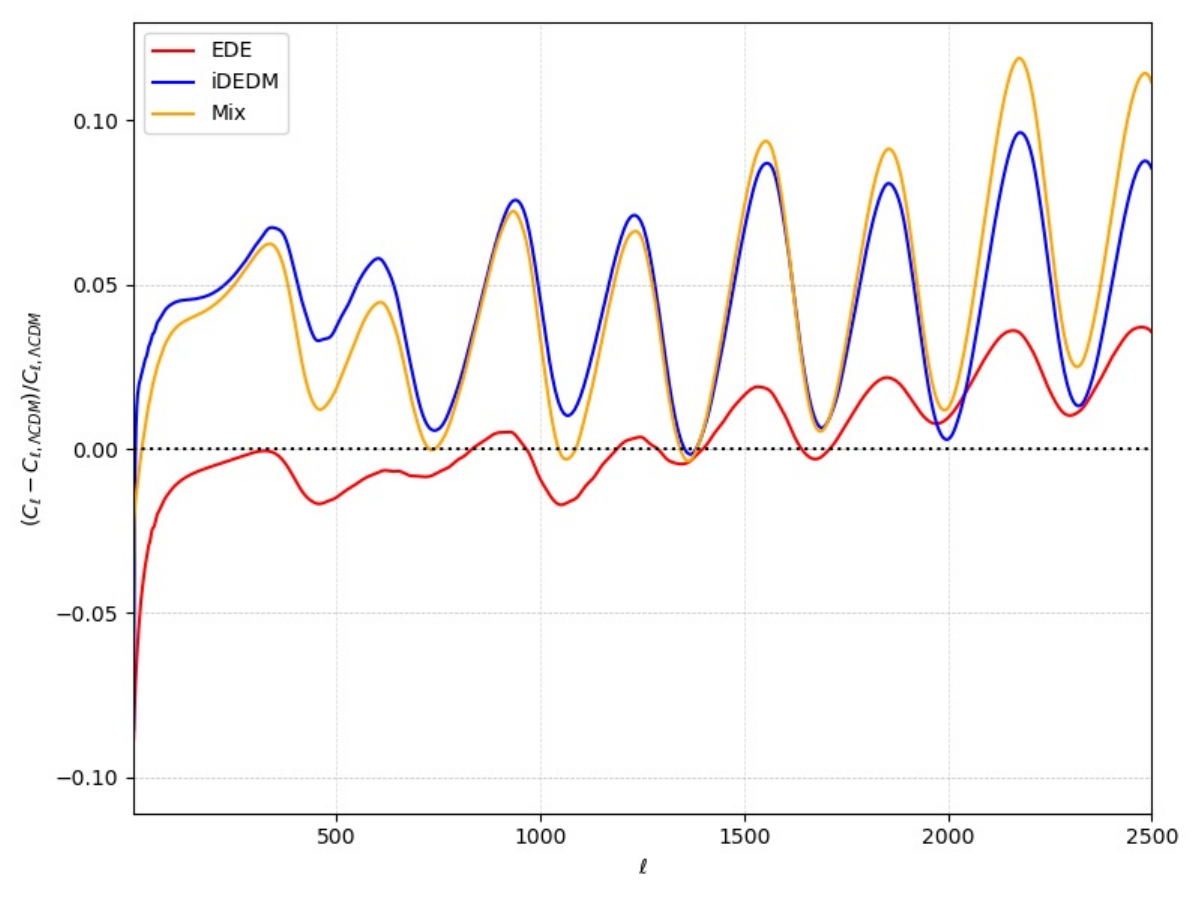}
    \caption{Relative difference in the CMB temperature power spectrum with respect to $\Lambda$CDM, $(C_\ell - C_\ell^{\Lambda \mathrm{CDM}})/C_\ell^{\Lambda \mathrm{CDM}}$.
    The spectra are computed using the best-fit parameters from each model.}
    \label{fig:Cl_sabun}
\end{figure}

While the statistical improvement is marginal, understanding the physical mechanisms driving this result provides further insight into the limitations of the mixed model.
Both EDE and iDEDM independently tend to favor a larger present-day cold dark matter density.
As mentioned in Sec.~\ref{sec:EDE_LSS}, in the EDE scenario, increasing $\omega_{\rm cdm}$ is necessary to maintain a good fit to the CMB power spectrum while raising $H_0$.
Meanwhile, in iDEDM, the present-day DM density increases to compensate for the reduced DM abundance at early times due to energy transfer from DE to DM.
Their combination naturally leads to a higher total matter density in the mixed model, which in turn reduces the angular diameter distance $D_A$.
As a result, the reduction in the sound horizon $r_s$ caused by EDE is no longer sufficient to maintain the observational value of $\theta_s=r_s/D_A$, leading to a mismatch in the angular scale of the first acoustic peak in the CMB spectrum.
In Fig.~\ref{fig:Cl_sabun}, we plot the relative deviation of the CMB temperature power spectrum, defined as
$\Delta C_\ell / C_\ell \equiv (C_\ell - C_\ell^{\Lambda \mathrm{CDM}})/C_\ell^{\Lambda \mathrm{CDM}}$.
This shows that the mixed model already deviates slightly from $\Lambda$CDM at $\ell \sim 200$, where the first peak is located.  
Increasing $\omega_{\rm cdm}$ — as required by both EDE and iDEDM — tends to enhance the amplitude of this peak and would further exacerbate the deviation.
This limitation restricts how much $\omega_{\rm cdm}$ can be increased in the mixed model, possibly explaining the suppression of the EDE contribution.

Indeed, this underscores the critical role that the total matter density $\Omega_m$ plays in the simultaneous resolution of the $H_0$ and $S_8$ tensions---a point also highlighted in Ref.~\cite{2407.01173toda}, where incompatible trends in $\Omega_m$ from early- and late-time components prevented a successful resolution.
As shown in Table~\ref{tab:result}, the 95\% upper bound on $f_{\rm EDE}$ decreases from $<0.128$ in the EDE-only case to $<0.113$ in the mixed model.
Moreover, the sound horizon $r_s$ slightly increases from $r_s = 143.673 \pm 1.709$~Mpc to $ r_s = 144.523\pm 1.950$~Mpc, further supporting the idea that EDE becomes less effective in modifying the early expansion history.
We also note that there is no significant degeneracy between $f_{\rm EDE}$ and $\xi$ in the mixed model, as shown in Fig.~\ref{fig:triangle}, which reinforces the interpretation that the suppression of EDE arises from physical constraints, rather than parameter degeneracies.

\begin{figure}[t]
    \centering
    \includegraphics[width=8cm]{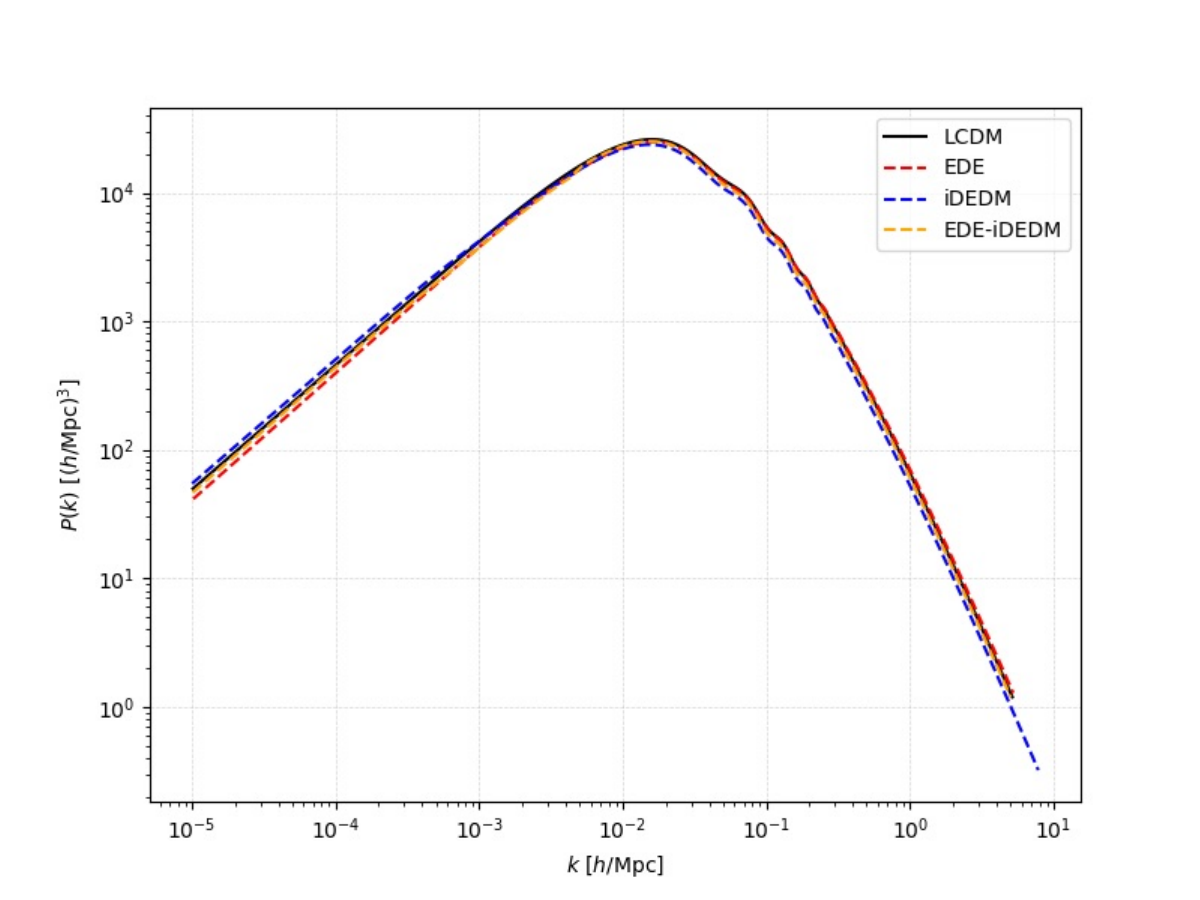}
    \caption{Matter power spectrum $P(k)$ for the $\Lambda$CDM, EDE-only, iDEDM-only, and mixed models, computed at $z=0$ using best-fit parameters.}
    \label{fig:mPS_hikaku}
\end{figure}

\begin{figure}
    \centering
    \includegraphics[width=8cm]{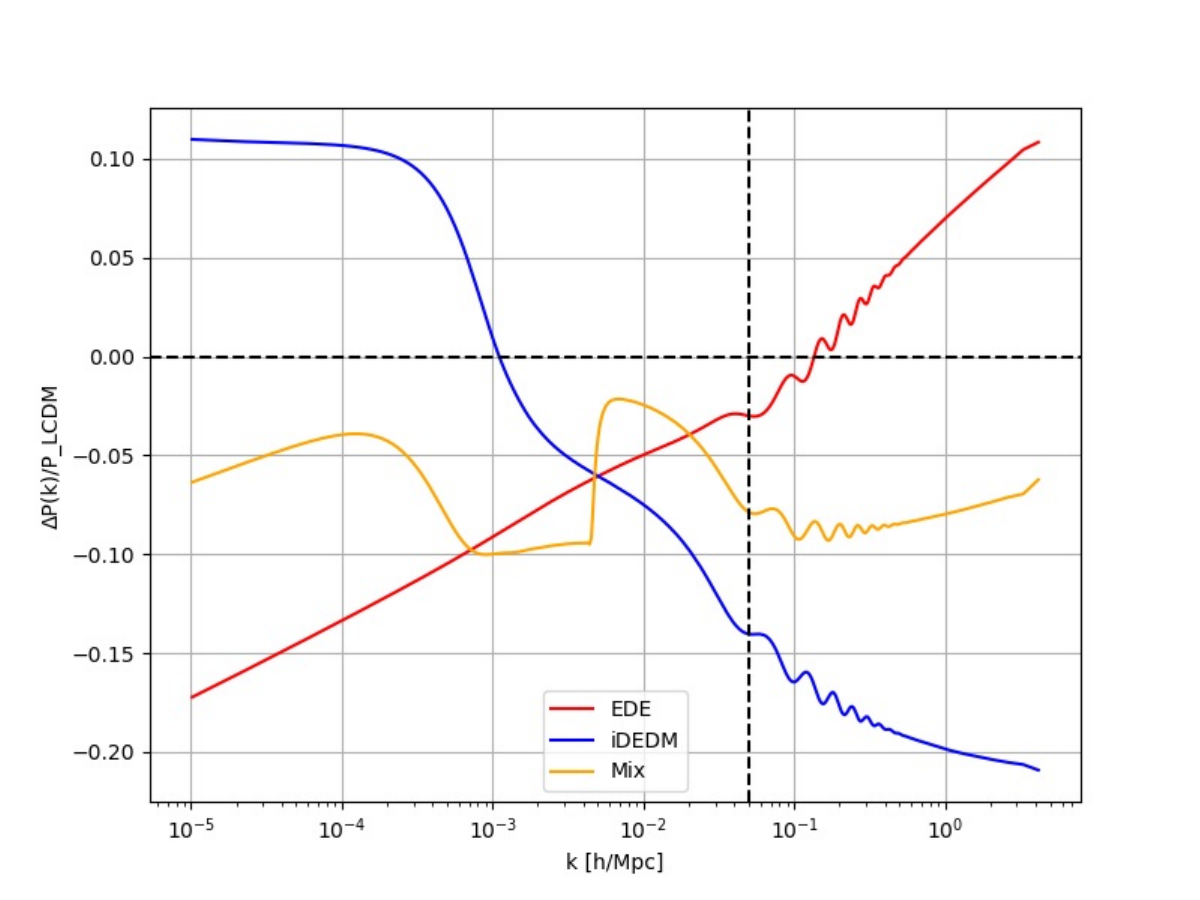}
    \caption{Relative difference in the matter power spectrum with respect to $\Lambda$CDM, defined as $\Delta P/P \equiv (P - P_{\Lambda \mathrm{CDM}})/P_{\Lambda \mathrm{CDM}}$.
    }
    \label{fig:mPS_sabun}
\end{figure}

We now examine the matter power spectrum to assess the scale-dependent impact of each model.
The matter power spectra $P(k)$ for all models are shown in Fig.~\ref{fig:mPS_hikaku}, and their relative deviations from $\Lambda$CDM are shown in Fig.~\ref{fig:mPS_sabun}, where $\Delta P/P \equiv (P - P_{\Lambda \mathrm{CDM}})/P_{\Lambda \mathrm{CDM}}$.
At small scales ($k \gtrsim 0.1~h/\mathrm{Mpc}$), the mixed model exhibits significant suppression in $P(k)$, similar to the iDEDM-only model.
This suppression arises primarily from the reduced DM density at early times due to the DE to DM energy transfer, which slows the growth of structure during the matter-dominated era and leads to a lower value of $\sigma_8$.
In contrast, the EDE-only model shows an enhancement of small-scale power, as the increase in $\omega_{\rm cdm}$ — required to maintain the fit to CMB while raising $H_0$ — boosts the amplitude of matter fluctuations.
The combined effect in the mixed model highlights the compensating role of iDEDM in offsetting the enhanced $\sigma_8$ induced by EDE.
At large scales ($k \lesssim 10^{-2}~h/\mathrm{Mpc}$), the mixed model does not exhibit the enhancement seen in the iDEDM-only case.
This suppression appears to result from the opposing effects of EDE and iDEDM on the early-time dark matter density:
while iDEDM reduces the DM abundance and delays the matter-radiation equality, EDE increases $\omega_{\rm cdm}$ to maintain the CMB fit, thereby raising the early-time DM density.
These competing effects partially cancel out, weakening the delayed-equality-induced growth enhancement from iDEDM and leading to a suppressed amplitude at large scales.
As seen in Fig.~\ref{fig:mPS_sabun}, the mixed model exhibits a mild bump in the linear matter power spectrum around $k \sim 0.01~h/\mathrm{Mpc}$, which is absent in both the EDE- and iDEDM-only cases.
This feature may result from a non-trivial interplay between EDE-induced changes in the sound horizon and iDEDM-driven modifications to the growth history.
Although the origin of this localized feature warrants further investigation, we have confirmed that it does not significantly affect integrated observables such as $\sigma_8$, nor does it conflict with current observational constraints.

Although our mixed model does not fully resolve either the $H_0$ or $S_8$ tension, these findings underscore both the potential and limitations of combining early- and late-time modifications in cosmological models\footnote{In the first version (v1) of this manuscript, we included a plot comparing the predicted $f\sigma_8(z)$ to DESI data points. However, we later realized that the data used were not based on published tabulated values. Accordingly, we have removed the figure and associated discussion in this version to ensure scientific accuracy. We apologize for the oversight.}.

\section{Conclusion}
\label{sec:conclusion}

In this work, we investigated whether a mixed cosmological model combining early dark energy (EDE) and interacting dark energy-dark matter (iDEDM) can simultaneously resolve the $H_0$ and $S_8$ tensions.
The EDE model raises $H_0$ by reducing the sound horizon $r_s$, but it also tends to increase the total matter density, thereby enhancing $S_8$ and exacerbating the tension with large-scale structure observations.
To address this issue, we incorporated the iDEDM model, in which energy transfer from dark energy to dark matter reduces the early-time DM density and suppresses the growth of structure, thereby lowering the amplitude of the matter power spectrum.

We performed a Markov Chain Monte Carlo analysis using a combination of data from Planck 2018, DESI-DR1, Pantheon+, SH0ES, and DES-Y1.
We found that the mixed model alleviates $S_8$ to a similar extent as the iDEDM-only model.
However, the contribution of EDE is reduced compared to its effect in the EDE-only case, as reflected in the smaller increase in $H_0$.
This suppression likely arises from the fact that both EDE and iDEDM individually prefer a large present-day DM density, resulting in a higher matter density in the mixed model, which in turn reduces the angular diameter distance.
As a result, the reduction in the sound horizon $r_s$ induced by EDE is no longer sufficient to maintain the angular scale of the sound horizon $\theta_s$, limiting the increase in $H_0$.
Therefore, while our mixed model helps alleviate both tensions, it cannot fully resolve them.

These results highlight the challenge of simultaneously addressing both the $H_0$ and $S_8$ tensions.
Even though the growth of structure can be effectively suppressed by mechanisms such as iDEDM, which operate across both early and late times, precise CMB measurements---especially the angular scale $\theta_s$---place strong constraints on the effectiveness of EDE.
This suggests that a full resolution of both tensions may require more flexible models, such as those featuring time-dependent interactions or non-trivial dark sector dynamics, or mechanisms of suppressing the enhanced early integrated Sachs-Wolfe (eISW) effect induced by additional early-time or interacting dark sector energy components.

Looking ahead, upcoming high-precision surveys---such as CMB-S4, LiteBIRD, and DESI full-shape analyses---will offer critical insights into both early- and late-time dynamics.  
These future datasets could provide stringent tests for combined models like ours, and may help distinguish between competing explanations for the tensions.
Moreover, EDE framework remains subject to the so-called coincidence problem---why EDE appears at the matter-radiation equality.
Furthermore, the recent release of the Planck NPIPE (PR4) data, which offers improved systematics particularly at low multipoles, provides an opportunity for future analyses to revisit our mixed model with potentially tighter constraints.
We leave such investigations to future work.

\section*{Acknowledgements}
This work was supported by the Scholarship Fund for Young/Women Researchers.

\bibliographystyle{apsrev4-2}
%

\end{document}